\documentclass[conference]{IEEEtran}
\IEEEoverridecommandlockouts

\usepackage{amsmath}
\usepackage{amssymb}
\usepackage{mathtools}
\usepackage{amsthm}

\usepackage[capitalize,noabbrev]{cleveref}

\usepackage{url}

\usepackage{setspace}
\usepackage{bm}
\usepackage{balance}
\usepackage{url}  
\usepackage{soul}
\usepackage{tikz}
\usepackage{color}
\usepackage{braket}
\usepackage{dsfont}
\usepackage{amsthm}
\usepackage{lipsum}
\usepackage{xcolor}         % colors
\usepackage{amsmath}
\usepackage{amssymb}
\usepackage{caption}
\usepackage{environ}
\usepackage{physics}
\usepackage{amsfonts}       % blackboard math symbols
\usepackage{booktabs}       % professional-quality tables
\usepackage{colortbl}
\usepackage{enumitem}
\usepackage{graphicx}
\usepackage{multirow}
\usepackage{nicefrac}       % compact symbols for 1/2, etc.
\usepackage{quantikz}
\usepackage{textcomp}
\usepackage{microtype}      % microtypography
\usepackage{subfigure}
\usepackage{tcolorbox}
\usepackage{subcaption}
\usepackage[T1]{fontenc}    % use 8-bit T1 fonts
\usepackage[utf8]{inputenc}% allow utf-8 input

\theoremstyle{plain}
\newtheorem{theorem}{Theorem}[section]

\newtheorem{lemma}[theorem]{Lemma}

\theoremstyle{definition}
\newtheorem{definition}[theorem]{Definition}

\theoremstyle{remark}
\newtheorem{remark}[theorem]{Remark}

\newcommand{\sol}{\textsc{HattriQ}}

\newcommand{\rev}[1]{\textcolor{black}{#1}}

\newcommand{\vect}[1]{\bm{#1}}

\definecolor{hadamard}{RGB}{0,191,175}
\definecolor{cnot}{RGB}{224,227,0}
\definecolor{rotation}{RGB}{11,244,70}
\definecolor{measurement}{RGB}{255,89,143}

\tikzset{
had/.style={draw, fill=hadamard!25},
cnot/.style={draw, fill=cnot!50},
rot/.style={draw, fill=rotation!25},
meas/.style={draw, fill=measurement!25}
}

\def\BibTeX{{\rm B\kern-.05em{\sc i\kern-.025em b}\kern-.08em
    T\kern-.1667em\lower.7ex\hbox{E}\kern-.125emX}}

\begin{document}

\date{}

% Leveraging Integrated Gradients for Feature Attribution in Quantum Circuits with the Design of \sol{}
\title{\sol{}: Designing Integrated Gradients for Feature Attribution in Quantum Machine Learning}
%\title{\sol{}: Input Attribution for Quantum Circuits using Integrated Gradients}

\author{
\IEEEauthorblockN{Nicholas S. DiBrita, Jason Han}
\IEEEauthorblockA{\textit{Rice University}\\ Houston, TX, USA}%\\ \texttt{nd52,jh146@rice.edu}}
\and
\IEEEauthorblockN{Younghyun Cho}
\IEEEauthorblockA{\textit{Santa Clara University}\\ Santa Clara, CA, USA}%\\ \texttt{younghyun.cho@scu.edu}}
\and
\IEEEauthorblockN{Hengrui Luo, Tirthak Patel}
\IEEEauthorblockA{\textit{Rice University}\\ Houston, TX, USA}%\\ \texttt{hl180,tp53@rice.edu}}
}

\maketitle

\begin{abstract}

Quantum machine learning (QML) algorithms have demonstrated early promise across hardware platforms, but remain difficult to interpret due to the inherent opacity of quantum state evolution. Widely used classical interpretability methods, such as integrated gradients and surrogate-based sensitivity analysis, are not directly compatible with quantum circuits due to measurement collapse and the exponential complexity of simulating state evolution. In this work, we introduce \sol{}, a general-purpose framework for computing amplitude-based input-attribution scores in circuit-based QML models. \sol{} supports the widely-used input amplitude embedding feature encoding scheme and uses a Hadamard test–based construction to compute input gradients directly on quantum hardware to compute integrated gradient attributions. We validate \sol{} on classification tasks across several datasets (Bars and Stripes, MNIST, FashionMNIST, and TFIM quantum data). %To our knowledge, \sol{} is the first quantum native gradient-based input attribution framework for QML models for amplitude embedding.

\end{abstract}

\section{Introduction}
\label{sec:intro}

%\TODO{Eveyrthing should be written in NeurIPS format (definition, remark, theorem, proof, etc.). I've copied the commands to do so from CircuitTree.}

Quantum machine learning (QML) uses quantum computing to enhance data analysis and pattern recognition in AI. By using quantum features like superposition and entanglement, QML algorithms have the potential to offer speedups over classical methods~\rev{\cite{biamonte_quantum_2017,derieux2024eqmarl,de2023makes}}. Current research emphasizes hybrid models, where quantum circuits work alongside classical optimizers~\cite{BhartiNISQalgo, cerezo2021variational}, with applications in classification, clustering, and generative tasks~\cite{preskill2018quantum,dibrita2024recon,zhang2023statistical,han2025enqode}. While limited by today’s hardware, QML holds promise for solving complex problems in fields such as healthcare, finance, and scientific computing as quantum systems advance\rev{~\cite{nicoli2023physics,hothem2024my,preskill2023quantum, cerezo2022challenges}}. Despite growing interest and experimental progress, QML models remain difficult to interpret due to the inherent opacity of quantum state evolution and the absence of intermediate observability mid computation~\cite{herbst2024exploring,pira2024interpretability,heese2025explaining}.

In classical machine learning, interpretability methods such as feature attribution play a critical role in understanding model predictions, particularly in sensitive and mission-critical domains like healthcare and autonomous systems~\cite{radenovic2022neural,zimmermann2023scale,agarwal2021neural,hooker2019benchmark,alvarez2018towards}. Attribution methods \cite{rudin2018stop,krishna2022disagreement} -- such as integrated gradients (IG)~\cite{Sundararajan_integrated_grad} -- assign importance scores to input features, revealing which aspects of the input most influence the model's output. These methods enhance transparency, support debugging, and build trust in model behavior. In contrast, existing QML pipelines provide little insight into how input features affect final measurement outcomes, especially when data is encoded and compressed into high-dimensional quantum state amplitudes~\cite{jerbi2021parametrized,bausch2020recurrent,preskill2018quantum,preskill2023quantum}.

%\footnote{\sol{} stands for \underline{H}adamard test-based input \underline{attri}bution score scheme for \underline{q}uantum models.}

We propose \sol{}, a methodology for computing input-attribution scores for quantum circuits. \sol{} adapts integrated gradients~\cite{Sundararajan_integrated_grad} to the quantum circuit setting, enabling attribution for amplitude embedding. Leveraging integrated gradients for quantum models is challenging, as larger models require working in exponentially large Hilbert spaces and manipulating complex amplitude vectors, making both analysis and simulation resource-intensive~\cite{xiongnode2ket,leineural}.

\begin{figure}[t]
    \centering
    \includegraphics[width=0.98\linewidth]{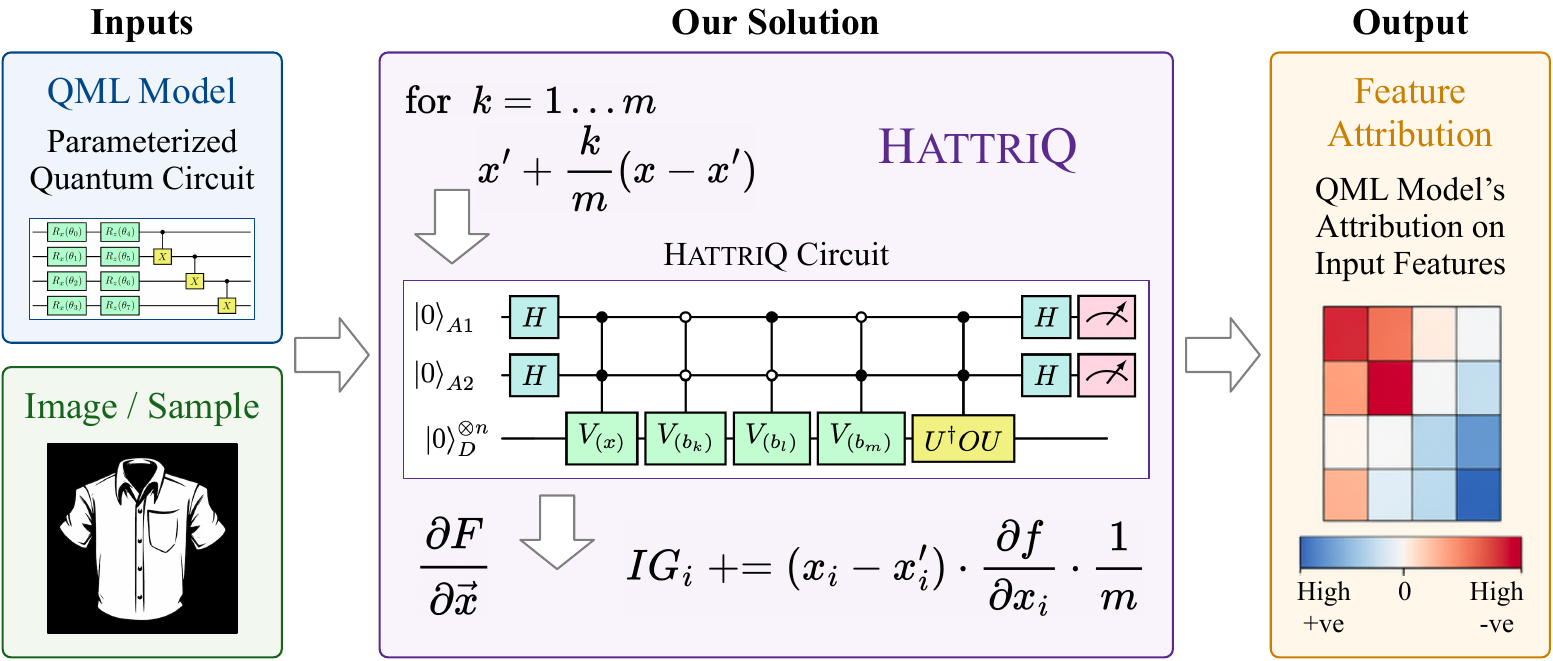}
    \caption{Overview of \sol{}'s execution flow for input feature attribution for a given QML model and image/sample.}
    \label{fig:overview}
    \vspace{-5mm}
\end{figure}

Another challenge is that quantum states are hidden from the user during computation. For large programs, we cannot simply record or log the hidden state after each circuit layer, as any attempt to measure the hidden state collapses the quantum state of the circuit entirely~\cite{gong2023learning,abbas2023quantum}; traditional (surrogate-based) sensitivity and Sobol/Shapley score methods~\cite{owen2014sobol,cho2025surrogate} cannot preserve unitarity in quantum circuits, making it difficult to understand how different signals are propagated through the circuit. To address this, \sol{} implements a Hadamard test–based construction that computes exact gradients directly on quantum hardware, without requiring access to internal quantum states (execution flow shown in Fig.~\ref{fig:overview}). For fault-tolerant quantum devices, where the impact of hardware noise~\cite{akhalwayatopological,wulearning,patelcurriculum} is negligible, we propose a parallelization mechanism to evaluate multiple gradient components concurrently. \sol{} enables input-level sensitivity analysis for amplitude-encoded data, a capability that classical IG cannot provide efficiently. Our main result provides a hardware-compatible way to obtain the necessary derivatives for IG calculation, without knowing the computer's internal quantum state.
Practically, this enables the identification of highly influential features and checks if these align with semantically meaningful regions in its prediction. It also allows us to compare encodings with similar accuracy but different attribution patterns, and reveal bias toward background or artifacts.

\vspace{1mm}

\noindent\textbf{Our contributions are as follows.}

\begin{itemize}
    \item We introduce a formalism to compute integrated gradients in QML models that use amplitude embedding for encoding data inputs.
    \item We present a quantum-native circuit construction based on the Hadamard test to compute exact feature gradients for amplitude-embedded input attribution.
    \item We provide a multi-ancilla-based parallelization technique that enables concurrent gradient computation on larger quantum devices with sufficient capacity.
    \item We evaluate \sol{} on classification tasks across Bars and Stripes~\cite{bowles2024better}, MNIST~\cite{lecun1998mnist}, and FashionMNIST~\cite{xiao2017fashion} datasets, demonstrating high-fidelity attribution. It is also evaluated under noisy hardware conditions.
    \item \sol{} efficacy is also demonstrated using a synthetic TFIM quantum dataset~\cite{franco_quantum_SHAP2026} to display its broad applicability under classical and quantum datasets.
    \item \sol{}'s code and dataset are open-sourced at:\\ \textit{\url{https://github.com/positivetechnologylab/HattriQ}}.
\end{itemize}

\section{Relevant Concepts}
\label{sec:background}

\subsection{Quantum States and Gates}

Quantum computations are performed by quantum circuits that manipulate qubits using logic gates. 
The \emph{state} of a qubit is represented as a vector: $\ket{\psi} = \beta_0\ket{0} + \beta_1\ket{1}$, where $\beta_i$ is a complex coefficient for basis state $\ket{i}$. The probability of measuring the qubit to be in state $\ket{i}$ is $|\beta_i|^2$, which means we must have $|\beta_0|^2 + |\beta_1|^2 = 1$~\cite{schuld2019quantum,silver2023mosaiq}.

For an $n$ qubit system, the statevector is a complex vector $\ket{\psi} \in \mathbb{C}^{2^n}$ that is normalized $\braket{\psi} = 1$. States are then written in terms of an orthonormal basis; the conventional choice is the computational basis. If we define $b_k$ as the bitstring corresponding to integer $k$, we can define the computational basis as the set $\{\ket{b_k} \, \forall \, k\in \mathbb{Z}, 0 \leq k \leq 2^n-1 \}$.  Our state can then be expressed as $\ket{\psi} = \sum_{k=0}^{2^n-1}\beta_k\ket{b_k}$~\cite{schuld2019quantum,silver2023mosaiq}. Logic gates are represented by unitary matrices ($U$) acting on states: $U\ket{\psi_1}=\ket{\psi_2}$. Circuits are constructed by composing sequences of gates together~\cite{white2001exploring,srinivasan2018learning}.

\subsection{Parameterized Quantum Circuits}

\rev{We study quantum models that can be represented by unitary circuits $U$ and measured observables $O$. While our technique is broadly applicable to general classes of QML models, for demonstration purposes, we focus on circuits with trainable gate parameters.} These parameterized quantum circuits (PQCs) are also referred to as variational quantum circuits and have found extensive applications in quantum machine learning, quantum chemistry, and other areas of quantum optimization~\cite{BhartiNISQalgo, cerezo2021variational}. Often, the trainable gates in PQCs are rotation gates that rotate the quantum state by an angle parameter. There are many possible ways to arrange a PQC; the fixed structure of a PQC is referred to as an ansatz, and is analogous to fixing a neural network architecture. Let $\vect{x} \in \mathbb{R}^D$ be a data point, and $ V(\vect{x})\ket{0} = \ket{x} \in \mathbb{C}^{2^n}$ be the quantum state that encodes it, with $V(\vect{x})$ being the circuit that performs the encoding. Let $U(\vect{ \theta})$ be a PQC with trainable parameters $\vect{ \theta }$~\cite{NEURIPSSchleich}, and $O$ be a Hermitian operator that represents the observable measured for the model output. We consider quantum models which apply some circuit operations to the input state $\ket{x}$ and then compute an expectation value, written as
\begin{equation}
\textstyle F(\vect{x} \,; \vect{\theta}) = \bra{x} U^{\dagger}(\vect{\theta}) \, O \,U (\vect{\theta}) \ket{x}.
\label{eq:model}
\end{equation}
In the more general case, we might compose $F(\vect{x} \,; \vect{\theta})$ with some other (likely nonlinear) function to add complexity to our model: our discussion generalizes simply by applying the chain rule~\cite{cerezo2021variational} in the gradient computation as introduced next.

\rev{The same is true for hybrid quantum-classical models, without substantial change to the methodology. In hybrid architectures, a quantum layer with amplitude encoding can be treated as a differentiable block: \sol{} supplies its input gradient, which can then be combined with classical IG in preceding layers via the chain rule. }
% QML models commonly make use of tanh, sigmoid, and ReLU functions for this purpose~\cite{XXX}.

\begin{remark}
Note: we do not place any specific requirements on $U$, except that it must be a valid unitary operator. In most applications, however, $U$ will have a fixed gate structure (ansatz). Some subset of these gates will depend on variational parameters $\vect{\theta}$, which are then optimized to minimize the loss. \rev{Later, we will also require that observable $O$ be unitary as well as Hermitian, as is the case with the standard Pauli operators.}
\end{remark}

\subsection{Integrated Gradients}

We base our technique on the integrated gradients method proposed in~\cite{Sundararajan_integrated_grad}. This work studies the problem of attributing the prediction of deep learning networks to input features in a sample. Integrate gradients benefit from an axiomatic formulation, with guarantees about their sensitivity and implementation invariance~\cite{Sundararajan_integrated_grad, mudrakartaModel}.

In addition to its superior theoretical properties, this method for attribution also only relies on a small number of model evaluations and gradient computations, without the need for additional knowledge of the hidden state~\cite{Sundararajan_integrated_grad}. This is highly desirable in the quantum setting, where measuring and storing the internal state at multiple points during the computation would incur significant overhead. 
\begin{definition}[Attribution Score]
The integrated gradients attribution of a sample $\vect{x}$ relative to baseline $\vect{x}'$ is given as the following integral:
\begin{equation}
IG_i(x) = (x_i - x_i') \, \int_0^1 \pdv{F(x' + \alpha \cdot (x - x'))}{x_i} d\alpha .
\end{equation}
The calculated value $IG_i$ is the integrated gradient attribution for the $i^{th}$ feature and represents the contribution of that feature to the final model prediction.
\end{definition}
%Generally for application purposes, this integral is discretized and evaluated numerically.

\section{Feature Gradients}
\label{sec:setup}

In this section, we introduce the most popular schemes for encoding data features into a quantum circuit calculation: (1) angle embedding and (2) amplitude embedding~\cite{havlivcek2019supervised,schuld2018supervised,lloyd2020quantum,iten2016quantum,schuld2019quantum}. For each of these encoding methods, we introduce our methodology for computing the gradients with respect to those encoded features, attributing the circuit output to features.

% Later on, we restrict $H$ to be some product of Pauli operators, but for most of the derivation, $H$ is allowed to be generic.

\subsection{Angle Embedding (or Encoding)}

For angle-embedded data, the preparation circuit $V(\vect{x})$ consists of rotation gates, $\{R(x_i)\}$ each of which depends on an angle parameter. The angle parameters used are the features $x_i$. In such cases, the gradient with respect to the features can be computed natively using the well-known parameter-shift rule \cite{MitaraiParamShift, SchuldParamShift}, which allows computing the gradient of quantum circuits by re-executing them with shifted parameter values. For a quantum gate parametrized by $\theta_i$ and with only two distinct eigenvalues $\pm r$, it has been shown~\cite{SchuldParamShift}:
\begin{equation}
    \textstyle\pdv{F}{\theta_i} = r\,[F(\theta_i + s) - F(\theta_i - s)]
    \label{eq:paramshift}
\end{equation}
where $s = \frac{\pi}{4r}$ is the required shift. While at first glance this formula is reminiscent of a standard finite difference, it differs in that the shift $s$ is not taken to be infinitesimal, and the result is exact. This requires two additional circuit evaluations per parameter, making the gradient calculation linear in the number of parameters. While Eq.~\ref{eq:paramshift} is not generally applicable to all gates, many parameterized gates, like single qubit rotations, do satisfy the eigenvalue requirements, and parameter shift has been utilized in a variety of quantum optimization settings~\cite{SchuldClassifiers, cerezo2021variational}. Additional rules have been formulated that generalize this result to additional kinds of parameterized gates~\cite{WierichsGenParamShift}.
% \TODO{Why not just use angle embedding if the gradient calculation is simple? Why do people bother to use amplitude embedding? What are the benefits of each?}

While its simplicity makes angle embedding an attractive choice for near-term applications, the number of encoded features typically grows only linearly with the number of qubits~\cite{SchuldClassifiers}, meaning the angle encoding does not make full use of the exponentially large Hilbert space, and does not reach the information upper bound on a sphere~\cite{luo2024spherical}.

\subsection{Amplitude Embedding (or Encoding)}

In the amplitude embedding case~\cite{khan2024beyond}, data features are encoded as amplitudes of the input state: $\ket{x} = \sum_i x_i \ket{b_i}$. Unlike the angle-embedding case, this allows for encoding exponentially many input features relative to the number of qubits, thereby expanding the circuit's information capacity. While it is generally true that the preparation circuit $V(\vect{x})$ is unitary, applying the parameter-shift rule for this purpose is not feasible because the circuit's dependence on the input features is complex. In particular, most state preparation circuits will have structures that change based on particular $\ket{x}$~\cite{buhrman2024state}, meaning any differentiation routine will necessarily depend on a complex and changing parameterization. Furthermore, there may be state preparation routines that do not satisfy the two-eigenvalue criteria mentioned above. In such cases, one would need to use the linear-combination-of-unitaries approach~\cite{SchuldParamShift}, which requires additional matrix decompositions and circuit evaluations. To address this challenge, we provide a novel circuit-based method for calculating input gradients that is independent of the routine used for $V(\vect{x})$.

\begin{lemma}[Input Gradient]
\label{lemma:gradientcircuit}
For the general case, assume the amplitudes of an amplitude-encoded input are complex valued, so that each $x_k = c_k + \mathbf{i}\,d_k$. Then, the input gradients with respect to the function given in Eq.~\ref{eq:model} are as follows for real-valued and complex-valued components.
\vspace{-2mm}
\begin{align*}
    \textstyle\pdv{F}{c_k} = 2 \Re[\bra{b_k}U^{\dagger}(\vect{\theta}) \, O \,U (\vect{\theta})\ket{x}]
    \\
    \textstyle\pdv{F}{d_k} = 2 \Im[\bra{b_k}U^{\dagger}(\vect{\theta}) \, O \,U (\vect{\theta})\ket{x}]
    \notag
\end{align*}
\end{lemma}
\vspace{-4mm}
\begin{proof}
The result is elegant to prove upon judicious use of the product rule for derivatives. For compactness, define $\Tilde{O} = U^{\dagger}(\vect{\theta}) \, O \,U (\vect{\theta})$ Then, after rewriting Eq.~\ref{eq:model}, we have: 
\begin{align*}
    F(\vect{x} \,; \vect{\theta}) &= \biggl(\sum_{i=0}^{2^n-1} \bra{b_i} x_i^* \biggl) \Tilde{O} \biggl(\sum_{j=0}^{2^n-1} x_j \ket{b_j}\biggl)\\& = \biggl(\sum_{i=0}^{2^n-1} \bra{b_i} (c_i -\mathbf{i}\,d_i) \biggl) \Tilde{O} \biggl(\sum_{j=0}^{2^n-1} (c_j + \mathbf{i}\,d_j) \ket{b_j} \biggl)\\
    &= \textstyle\sum_{i,j} (c_i -\mathbf{i}\,d_i) (c_j + \mathbf{i}\,d_j) \bra{b_i} \Tilde{O} \ket{b_j}
\end{align*}
Taking the derivative with respect to $c_k$:
\begin{align*}
    \textstyle\pdv{F}{c_k}  &= \textstyle\sum_{ij} \textstyle\pdv{c_i}{c_k} \left( c_j + \mathbf{i}\,d_j \right) \bra{b_i}\Tilde{O}\ket{b_j} + (c_i -\mathbf{i}\,d_i)\textstyle\pdv{c_j}{c_k} (c_i -\mathbf{i}\,d_i) \bra{b_i}\Tilde{O}\ket{b_j}\\
    &= \textstyle\sum_{ij} \delta_{ik} \left( c_j + \mathbf{i}\,d_j \right) \bra{b_i}\Tilde{O}\ket{b_j} + \delta_{jk} (c_i -\mathbf{i}\,d_i) \bra{b_i}\Tilde{O}\ket{b_j}\\
    &= \textstyle\sum_{j} \left( c_j + \mathbf{i}\,d_j \right) \bra{b_k}\Tilde{O}\ket{b_j} + \sum_{i} (c_i -\mathbf{i}\,d_i) \, \bra{b_i}\Tilde{O}\ket{b_k}\\
    &= \textstyle\sum_{j} \left( c_j + \mathbf{i}\,d_j \right) \bra{b_k}\Tilde{O}\ket{b_j} + (c_j -\mathbf{i}\,d_j) \, \bra{b_j}\Tilde{O}\ket{b_k}\\
    &= \textstyle\sum_{j} 2\,\Re[\left( c_j + \mathbf{i}\,d_j \right) \bra{b_k}\Tilde{O}\ket{b_j}] \\&= 2 \Re [ \bra{b_k}\Tilde{O}\textstyle\sum_{j}\left( c_j + \mathbf{i}\,d_j \right) \ket{b_j}] \\
    &= 2 \Re[\bra{b_k}\Tilde{O}\ket{x}] = 2 \Re [\, \bra{b_k}U^{\dagger}(\vect{\theta}) \, O \,U (\vect{\theta})\ket{x}\, ]
\end{align*}
Here, $\delta_{ik}$ is the Kronecker delta, and we have made use of the fact that $\Tilde{O}^\dagger = \Tilde{O}$. A similar derivation exists for $\pdv{F}{d_k}$. We exclude it here for brevity.
\end{proof}

Lemma~\ref{lemma:gradientcircuit} gives a compact expression for the $k^{th}$ component of the gradient in terms of the trained model circuit $U(\vect{\theta})$, its Hermitian conjugate $U^\dagger(\vect{\theta})$, Hermitian observable $O$, and amplitude embedded state $\ket{x}$. In this work, we are primarily concerned with the case where all amplitudes are real, $x_i = c_i$, as this is the most common setting encountered in classical data analysis.

\begin{remark}
If we add the constraint that $O$ be unitary as well as Hermitian, then $U^\dagger(\vect{\theta})OU(\vect{\theta})$ corresponds to a valid quantum circuit. The obvious choices for $O$ that satisfy this are Pauli operators or strings of Pauli operators~\cite{dion2024efficiently}, which are available on most devices as both measurement and gate operations.
\end{remark}
\section{Calculating on Quantum Hardware}
\label{sec:design}

\subsection{Hadamard Test}
\label{sec:hadamardtest}

\begin{definition}[Hadamard Test]
\label{def:Hadamardtest}
Given unitary operators $A$ and $B$ such that $A\ket{0}=\ket{a}$ and $B\ket{0}=\ket{b}$, the Hadamard test~\cite{montanaro2013survey,audenaert2008asymptotic,aharonov2006polynomial} is a method for encoding the value $\Re[\braket{a}{b}]$ into the expectation value of a quantum circuit observable. This is achieved by the following circuit:
\vspace{-2mm}
{
\begin{center}
\begin{quantikz}[row sep=2mm]
\lstick{$\ket{0}_0$} & \gate[style=had]{H} & \ctrl{1} & \octrl{1} & \gate[style=had]{H} & \meter[style=meas]{} \\
\lstick{$\ket{0}_1$} & \qw  & \gate[style=cnot]{A} & \gate[style=rot]{B}   & \qw   & \qw
\end{quantikz}
\end{center}}
The circuit for computing $\Im[\braket{a}{b}]$ is the same, with the addition of an $S^\dagger$ gate after the first $H$ gate.
\end{definition}
After the initial Hadamard gate $H = \frac{1}{\sqrt{2}}\big(\begin{smallmatrix}
  1 & 1\\
  1 & -1
\end{smallmatrix}\big)$, we have the state $\frac{1}{\sqrt{2}}[\ket{0}_0+\ket{1}_0]\ket{0}_1$. Applying A conditioned on 1 and B conditioned on 0 gives the entangled state 
\begin{align*}
    \textstyle \frac{1}{\sqrt{2}}[\,\ket{0}_0 B\ket{0}+\ket{1}_0 A\ket{0}\,] = \frac{1}{\sqrt{2}}[\,\ket{0}_0 \ket{b}+\ket{1}_0 \ket{a}\,].
\end{align*} 
The final Hadamard gate gives us 
\begin{align*}
    \textstyle \frac{1}{\sqrt{2}} \, ( \frac{1}{\sqrt{2}} (\ket{0}_0 + \ket{1}_1)\ket{b}_1 + \frac{1}{\sqrt{2}} (\ket{0}_0 - \ket{1}_1)\ket{a}_1\\
    = \frac{1}{2} ( \ket{0}(\ket{b}+\ket{a}) + \ket{1}(\ket{b}-\ket{a})).
\end{align*}
From this we can compute the probability of measuring qubit 0 to be 0 as 
\begin{align*}
    \textstyle P(0) = \frac{1}{2}[\,\bra{b}+\bra{a}) \cdot  \frac{1}{2}(\ket{b}+\ket{a}\,]\\
    = \frac{1}{4}[\,\braket{b}{b}+ \braket{b}{a}+ \braket{a}{b} +\braket{a}{a}\,] = \frac{1}{2}[\,1 + \Re[\braket{a}{b}]\,].
\end{align*}
This allows us to estimate the desired inner product by sampling from the probability distribution of additional qubits entangled with the system~\cite{SchuldParamShift}\footnote{\sol{} stands for \underline{H}adamard test-based input \underline{attri}bution score scheme for \underline{q}uantum models.}. Hadamard tests have been used previously to compute certain kinds of parameter gradients~\cite{BhartiNISQalgo, SchuldParamShift}, but not for feature gradients. We reframe our formulation in Lemma~\ref{lemma:gradientcircuit} to enable hardware-native gradient computation for input amplitudes.

\subsection{Gradient Calculation for Input Attribution}
\label{sec:gradforinputattr}

Lemma~\ref{lemma:gradientcircuit} and Definition~\ref{def:Hadamardtest} imply that we can calculate the feature gradient of a quantum model using circuit evaluations. We propose a circuit based on a Hadamard test:
\vspace{-2mm}
{
\begin{center}
\begin{quantikz}[column sep=1.5mm,row sep=2mm]
\lstick{$\ket{0}_A$ \,} & \gate[style=had]{H} & \ctrl{1} & \octrl{1} & \qw & \ctrl{1} & \qw & \gate[style=had]{H} & \meter[style=meas]{} \\
\lstick{$\ket{0}^{\otimes n}_D$} & \qw  & \gate[style=cnot]{V(x)} & \gate[style=rot]{V(b_k)}  & \gate[style=cnot]{U} & \gate[style=cnot]{O} & \gate[style=cnot]{U^\dagger} & \qw   & \qw
\end{quantikz}
\end{center}}

Here, $H$ is the aforementioned Hadamard gate, used to create equal superposition states. $V(x)$ is the preparation circuit that prepares $\ket{x}$. Similarly, $V(b_k)$ prepares the computational basis state $\ket{b_k}$. The wires that extend from one qubit register to another indicate control operations; these are multiqubit operations in which the state of one or more qubits in a target register undergoes a transformation, predicated on the state of the control register. In the above circuit, the control is always the ancilla qubit (indexed by $A$), while the targets are always the data qubits (indexed by $D$). The target state that triggers the control operation is indicated by the circle: filled circles indicate control gates triggered by the $\ket{1}_A$ state, while empty circles indicate control gates triggered by $\ket{0}_A$. As an example, consider the first controlled gate, controlled $V(\vect{x})$. This gate prepares $\ket{x}$ on the data register $D$ when the ancilla $A$ $\ket{1}$, and does nothing when $A$ is in the $\ket{0}$ state. 

The key feature of our technique is that gradients are computed from expectation values. This direct expectation-value measurement avoids the resource-intensive process of hidden-state tomography or classical simulation.
In essence, the difficulty in implementing the technique centers on the controlled operations, which, for a given architecture, might be designed or chosen to be compiled or approximated more efficiently than general multi-controlled gates, especially in hardware that supports native multi-qubit interactions. While it is true that multicontrol gates do incur high overhead on low connectivity platforms like superconducting hardware, it is possible that this overhead might be avoidable on other hardware platforms, like neutral atoms~\cite{delakouras2025multi, levineMulti} and trapped ions~\cite{trappedIonToffoli, trappedIonNbody}, where native multi-qubit operations are possible. For demonstration purposes, we focus on layers of CNOTs due to their prevalence in the literature; however, it is entirely possible to substitute this for another type of entangling gate, based on these native hardware constraints.

\begin{theorem}[]
\label{theorem:htest}
The above circuit returns the $k^{th}$ element of the gradient provided in Lemma~\ref{lemma:gradientcircuit}, encoded as the probability that qubit $A$ is measured as 0.
\end{theorem}

\begin{proof} The result can be seen almost directly from considering definition~\ref{def:Hadamardtest}. We use the subscript $A$ for the state of ancilla qubit(s), and the subscript $D$ for the state of data qubit(s). The final state of the circuit from Section~\ref{sec:gradforinputattr} before measurement is given by:
\begin{align*}
\ket{\psi} =& \left( I \otimes H \right) \cdot C\Tilde{O} \cdot \Bar{C}V(b_k) \cdot CV(x) \\&\cdot \left( I \otimes H \right) \cdot ( \ket{0}^{\otimes n}_D \otimes \ket{0}_A ) \\
=& \left( I \otimes H \right) \cdot C\Tilde{O} \cdot \Bar{C}V(b_k) \cdot CV(x) \\&\cdot \textstyle\frac{1}{\sqrt{2}} \left( \ket{0}^{\otimes n}_D \otimes \ket{0}_A +  \ket{0}^{\otimes n}_D \otimes \ket{1}_A\right) \\
=& \left( I \otimes H \right) \cdot \textstyle\frac{1}{\sqrt{2}} \left(  \ket{b_k}_D \otimes \ket{0}_A + \Tilde{O} \ket{x}_D \otimes \ket{1}_A \right) \\
=& \textstyle\frac{1}{2} \left( \ket{b_k}_D \otimes ( \ket{0}_A + \ket{1}_A) + \Tilde{O} \ket{x}_D \otimes (\ket{0}_A - \ket{1}_A) \right) \\
=& \textstyle\frac{1}{2} \left( (\ket{b_k} + \Tilde{O} \ket{x} )_D \otimes \ket{0}_A +  (\ket{b_k} - \Tilde{O}\ket{x})_D \otimes \ket{1}_A \right)
\end{align*}
Here, $C$ denotes the control operations that trigger on $\ket{1}_A$ and $\Bar{C}$ denotes the control operations that trigger on $\ket{0}_A$.  We also remark that $C\Tilde{O}$ can be decomposed into $U ^\dagger \, CO \, U$, due to the fact that $U^\dagger$ uncomputes $U$ whenever $CO$ is not triggered. From here, using the standard probability rule, we see that
\begin{align*}
P(A=0) =& |\textstyle\frac{1}{2}(\ket{b_k} + U^\dagger OU \ket{x} )|^2 \\&= \textstyle\frac{1}{4}|\braket{b_k}{b_k} + \braket{x} + \bra{b_k}U^\dagger OU\ket{x} \\&+ \bra{x} U^\dagger OU \ket{b_k}| \\
=& \textstyle\frac{1}{2}(1 + \Re[\,\bra{b_k}U^\dagger OU\ket{x}\,])
\end{align*}

The resulting measurement probability on the $A$ register is $P(A=0) = \textstyle\frac{1}{2}(1 + \Re[\,\bra{b_k}U^\dagger OU\ket{x}\,])$. Comparing with Lemma~\ref{lemma:gradientcircuit}, we see that the $k^{th}$ entry of the gradient is contained within the probability of measuring the ancilla to be 0. We can repeat this for each component. For a fixed number of measurement shots, this gives a linear relationship with the number of input features, i.e., one circuit is required per input feature, when a single ancilla qubit is used.
\end{proof}

\begin{table*}[t]
\centering
\caption{Datasets and models used for \sol{}'s evaluation, including the accuracies achieved.}
\label{tab:models}
\vspace{-1mm}
\scalebox{0.9}{
\renewcommand{\arraystretch}{1.3}{
\begin{tabular}{p{1.1cm}p{4.7cm}lllp{2.7cm}p{2.9cm}}
\toprule
\multirow{2}{*}{\textbf{Dataset}} & \multirow{2}{*}{\textbf{Binary Classes}} & \multirow{2}{*}{\textbf{Encoding}} & \multicolumn{2}{c}{\textbf{Circuit Structure}} & \multicolumn{2}{c}{\textbf{Accuracy (\%)}} \\ \cline{4-7} 
& & & \textbf{\# Qubits} & \textbf{\# Layers} & \textbf{Training} & \textbf{Testing} \\
\midrule
\multirow{2}{1.1cm}{\textit{Bars \& Stripes}} & (Bars, Stripes) & Amplitude & 4 & 8 & 96 & 95  \\ \cline{3-7}  &  & Angle & 8 & 8 & 95 & 95  \\
\midrule
\textit{NIST} & (0, 1), (3, 4), (5, 6), (6, 9), (1, 7) & Amplitude & 6 & 12 & 98, 100, 98, 96, 93 & 99, 100, 100, 98, 88  \\
\midrule
\textit{MNIST} & (0, 1), (3, 4), (5, 6), (6, 9), (1, 7) & Amplitude & 10 & 12 & 92, 88, 87, 62, 87 & 91, 82, 87, 68, 83 \\
\midrule
\textit{Fashion MNIST} & (Dress, Shirt), (Boot, Trousers), (Coat, Sandal), (Bag, Sandal), (Boot, Dress) & Amplitude & 10  & 10  & 74, 100, 96, 74, 90 & 70, 99, 95, 69, 91 \\
\midrule
\textit{TFIM} & (ordered, disordered) & Amplitude & 4 & 4 & 98 & 98 \\ 
\bottomrule
\end{tabular}}}
\end{table*}

\subsection{Gradient Calculation Parallelization}

We can further parallelize the component operations (the $k$'s in Theorem~\ref{theorem:htest}) by increasing the number of ancilla qubits. Doing so recovers the exponential scaling of input features with (ancilla) qubit count. For instance, if three components need to be executed in parallel, the following circuit can compute the $k^{th}$, $l^{th}$, and $m^{th}$ components concurrently.
\vspace{-1mm}
\begin{center}
\scalebox{0.87}{
\begin{quantikz}[column sep=1mm,row sep=2mm]
\lstick{$\ket{0}_{A1}$ \,} & \gate[style=had]{H} & \ctrl{1} & \octrl{1} & \ctrl{1} & \octrl{1} &  \qw & \ctrl{1} &  \qw & \gate[style=had]{H} & \meter[style=meas]{} \\
\lstick{$\ket{0}_{A2}$ \,} & \gate[style=had]{H} & \ctrl{1} & \octrl{1} & \octrl{1} & \ctrl{1}  & \qw & \ctrl{1} & \qw & \gate[style=had]{H} & \meter[style=meas]{} \\
\lstick{$\ket{0}^{\otimes n}_D$} & \qw  & \gate[style=cnot]{V_{(x)}} & \gate[style=rot]{V_{(b_k)}}  & \gate[style=rot]{V_{(b_l)}} & \gate[style=rot]{V_{(b_m)}}  & \gate[style=cnot]{U} & \gate[style=cnot]{O} & \gate[style=cnot]{U^\dagger} & \qw   & \qw
\end{quantikz}}
\end{center}

The circuit uses two ancilla qubits instead of one and measures both the ancilla simultaneously. A similar calculation as provided in Theorem~\ref{theorem:htest} gives an output probability of $P(A_1A_2=00) = \frac{1}{16}\big[ 4 + 2\Re [\bra{b_k}\Tilde{O}\ket{x} + \bra{b_l}\Tilde{O}\ket{x} + \bra{b_m}\Tilde{O}\ket{x}]\big]$. Similar expressions exist for $P(A_1A_2=01)$, $P(A_1A_2=10)$, and $P(A_1A_2=11)$, from which we can compute the expectation values needed.

The number of expectation values extracted from the system equals the number of encoded basis states, meaning it scales exponentially with the number of ancilla qubits. For $m$ ancilla qubits, we are able to compute $2^m - 1$ gradient components, preserving the exponential space savings associated with amplitude embedding. The general formula for computing the expectation values $ \forall k\neq r$ is \begin{equation}
\textstyle \Re[\bra{b_k}\Tilde{O}\ket{x}]
=2^{m-1}\sum_a (-1)^{a\cdot(k\oplus r)}
\Big(p_a-\frac{1}{2^m}\Big)
,
\end{equation} where $r=1^m$ is the all ones bitstring, which is reserved to prepare the state $\Tilde{O}\ket{x}$. We also have $p_a = P(A_{m-1}A_{m-2}...A_{0})$ as the probability of ancilla bitstring $a$. Here, the $\oplus$ and $\cdot$ in the exponent denote bitwise XOR and AND, respectively. To formalize this, we state the following theorem:

\begin{figure*}[t]
    \centering
    \subfigure[0 / 1]{\includegraphics[width=0.18\linewidth]{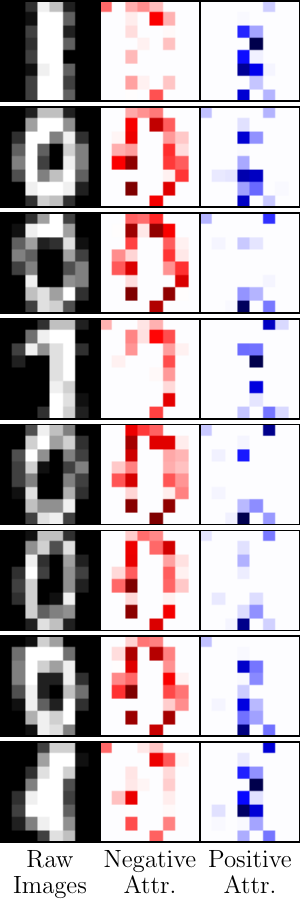}}
    \hspace{0.01\linewidth} % Space between subfigures
    \subfigure[3 / 4]{\includegraphics[width=0.18\linewidth]{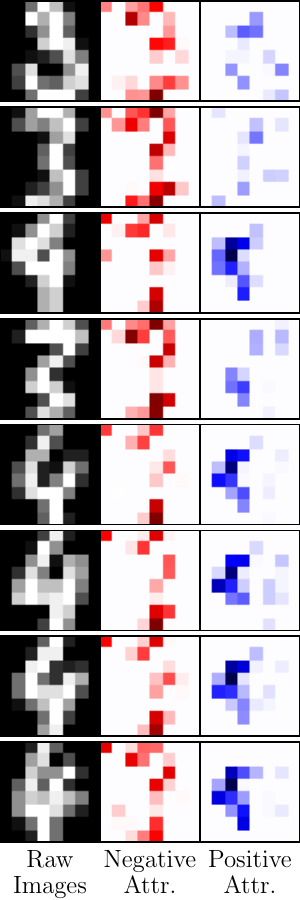}}
    \hspace{0.01\linewidth} % Space between subfigures
    \subfigure[5 / 6]{\includegraphics[width=0.18\linewidth]{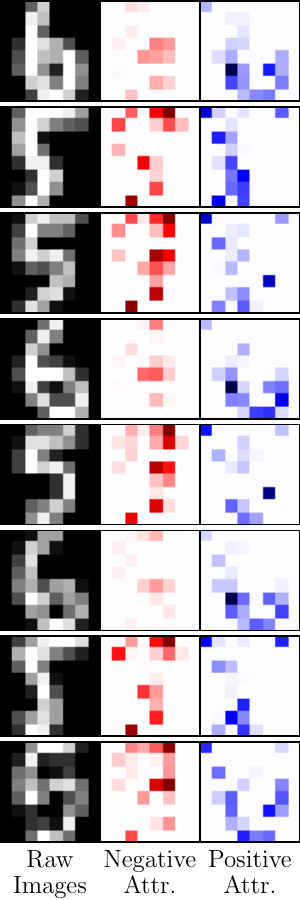}}
    \hspace{0.01\linewidth} % Space between subfigures
    \subfigure[6 / 9]{\includegraphics[width=0.18\linewidth]{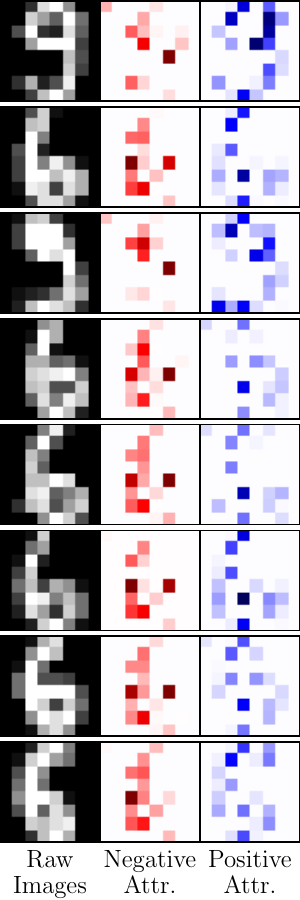}}
    \hspace{0.01\linewidth} % Space between subfigures
    \subfigure[1 / 7]{\includegraphics[width=0.18\linewidth]{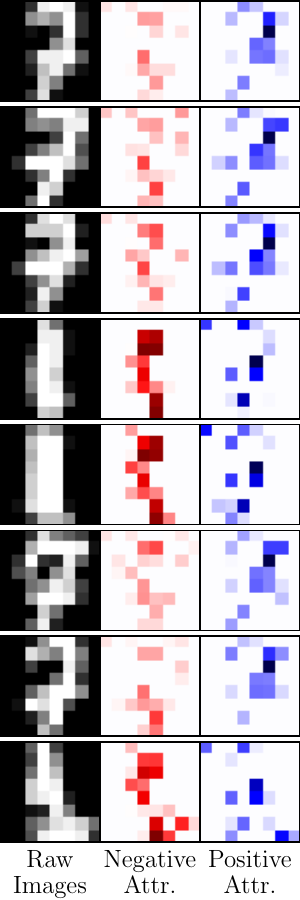}}
    \vspace{-3mm}
    \caption{Sample images and the accompanying integrated gradients attribution for various samples from the NIST dataset. Quantum models were trained for binary classification. Blue indicates positive attribution, red indicates negative attribution, and white indicates neutral attribution. We see patches and patterns of strong attributions for the trained classifier models.}
    \label{fig:NISTattributions}
    \vspace{-3mm}
\end{figure*}

\begin{figure*}[t]
    \centering
    \subfigure[Dress / Shirt]{\includegraphics[width=0.18\linewidth]{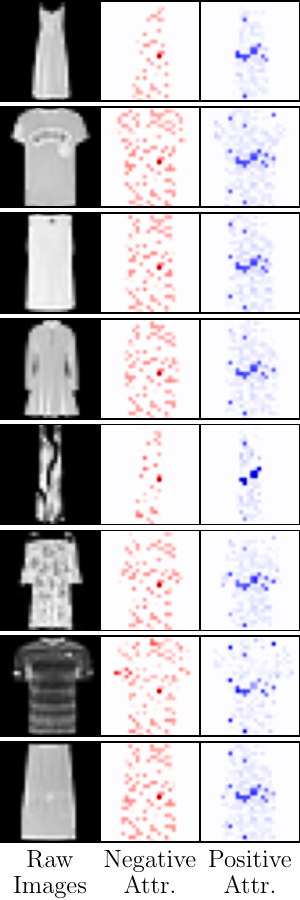}}
    \hspace{0.01\linewidth} % Space between subfigures
    \subfigure[Boot / Trousers]{\includegraphics[width=0.18\linewidth]{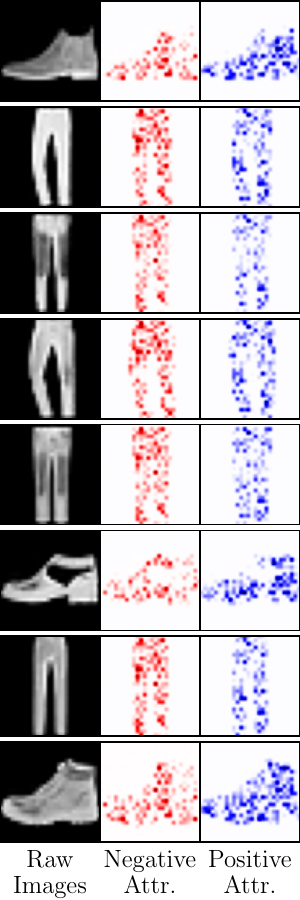}}
    \hspace{0.01\linewidth} % Space between subfigures
    \subfigure[Coat / Sandal]{\includegraphics[width=0.18\linewidth]{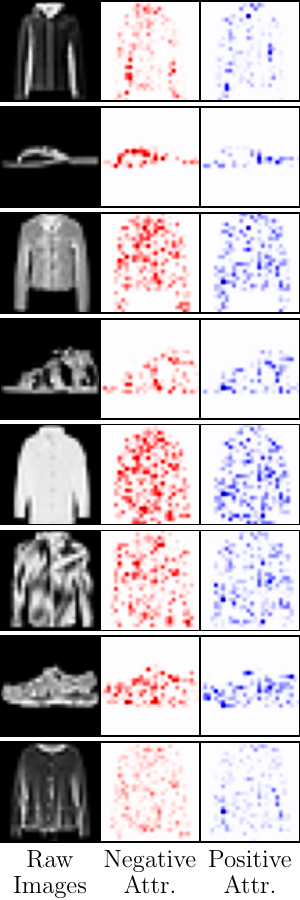}}
    \hspace{0.01\linewidth} % Space between subfigures
    \subfigure[Bag / Sandal]{\includegraphics[width=0.18\linewidth]{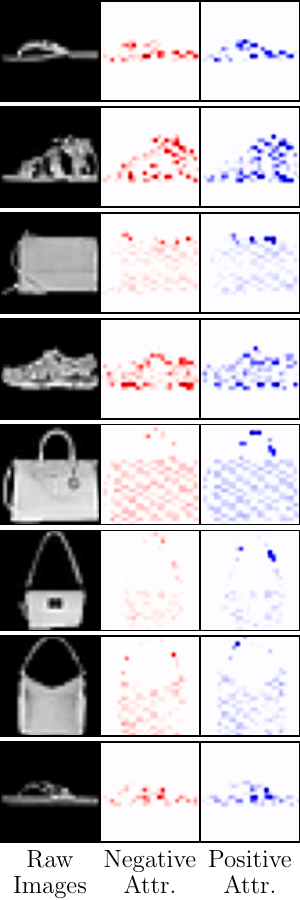}}
    \hspace{0.01\linewidth} % Space between subfigures
    \subfigure[Boots / Dress]{\includegraphics[width=0.18\linewidth]{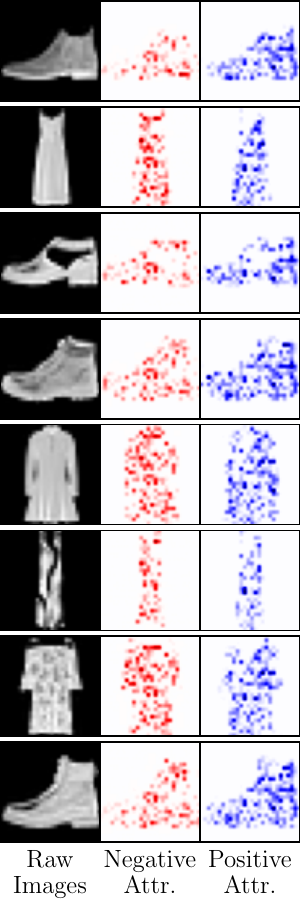}}
    \vspace{-3mm}
    \caption{\rev{Sample images and attributions for the FashionMNIST dataset using amplitude encoding.}}
    \label{fig:FashionAttribution}
    \vspace{-5mm}
\end{figure*}

\begin{figure}[t]
    \centering
    \subfigure[Angle Encoding]{\includegraphics[width=0.99\linewidth]{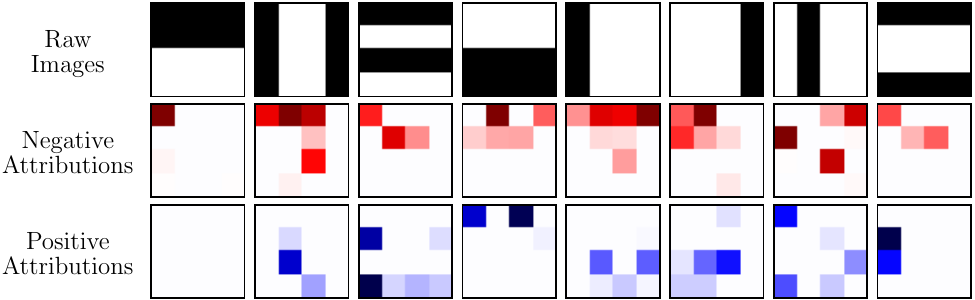}}
    \hfill
    \subfigure[Amplitude Encoding]{\includegraphics[width=0.99\linewidth]{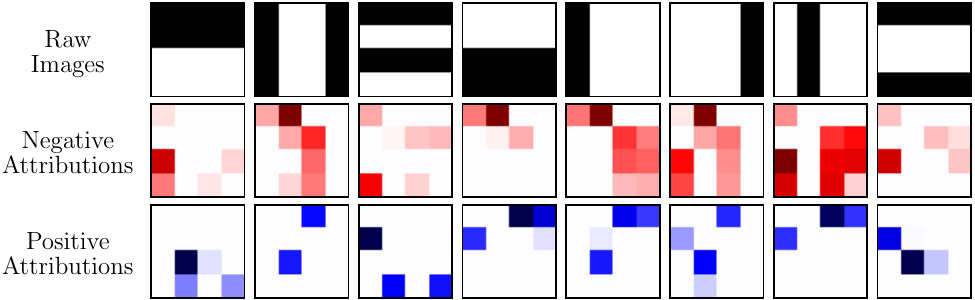}}
    \vspace{-3mm}
    \caption{Sample images and the accompanying integrated gradients attribution for the Bars and Stripes dataset for models using (a) angle encoding and (b) amplitude encoding.}
    \label{fig:BarsAndStripesAttr}
    \vspace{-5mm}
\end{figure}

\begin{figure*}[t]
    \centering
    \subfigure[0 / 1]{\includegraphics[width=0.3\linewidth]{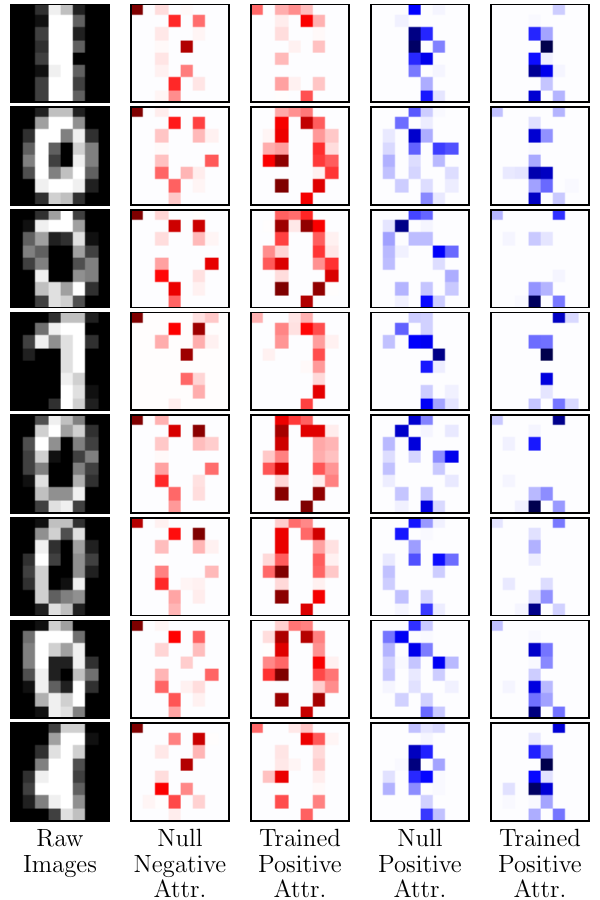}}
    \hspace{0.01\linewidth}
    \subfigure[3 / 4]{\includegraphics[width=0.3\linewidth]{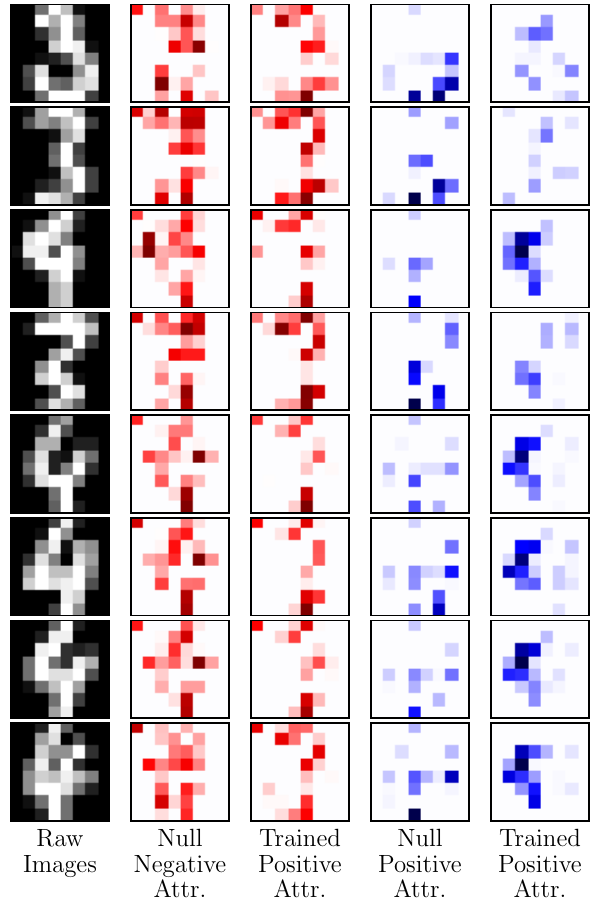}}
    \hspace{0.01\linewidth}
    \subfigure[5 / 6]{\includegraphics[width=0.3\linewidth]{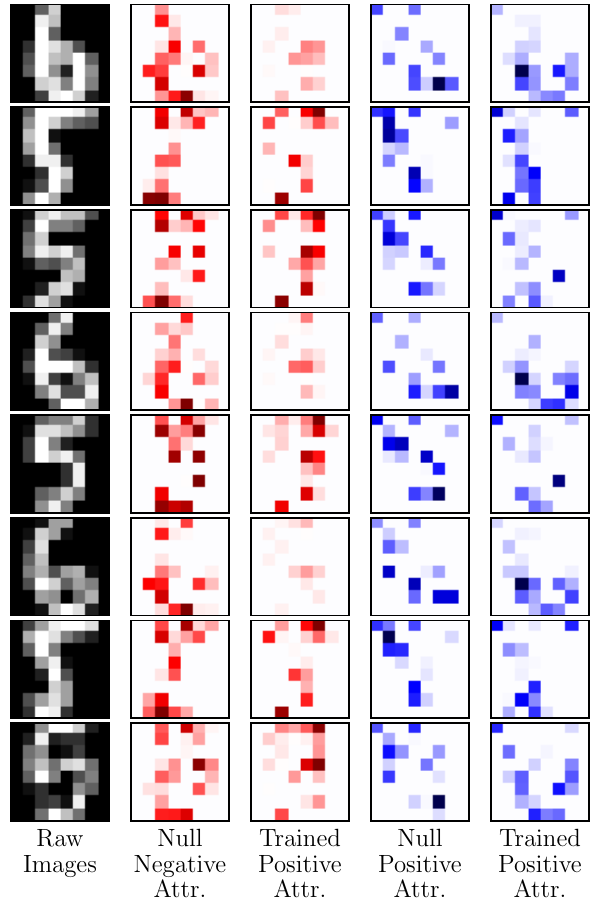}}
    \hspace{0.01\linewidth}
    \vspace{-3mm}
    \caption{Sample images and the accompanying integrated gradients attribution for various samples from the NIST dataset. Attributions are for untrained null models with parameters sampled from a uniform distribution on the interval $[0,\pi)$. For comparison, we re-plot attributions for trained models from Fig.~\ref{fig:NISTattributions} alongside the null model attributions. Blue (red) indicates positive (negative) attribution. We see from the lack of concentration that the null models fail to identify key features.}
    \vspace{-5mm}
    \label{fig:nulluniform}
\end{figure*}

\begin{theorem}[Parallel Hadamard-test with $m$ ancillas]
Let $D$ be an $n$-qubit data register and $A$ an $m$-qubit ancilla register.
Fix a \emph{reserved} ancilla string $r:=1^m$.
For every $s\in\{0,1\}^m\setminus\{r\}$, let $\ket{b_s}$ be an orthonormal
computational-basis state for $D$.
Let $\ket{x}$ be an arbitrary normalized data state and let $\tilde O$ be unitary.
Define
\[
r_s := \Re[\bra{b_s}\Tilde{ O}\ket{x}]\quad (s\neq r).
\]
Then a single circuit using $m$ ancillas yields all $2^m-1$ values $r_s$
(and hence all gradients $g_s=2r_s$) by measuring $A$.
\end{theorem}
\begin{proof}
Start in $\ket{0^m}_A\ket{0^n}_D$ and apply $H^{\otimes m}$ on $A$:
\[
\ket{\Psi_0}
=\frac{1}{\sqrt{2^m}}\sum_{s\in\{0,1\}^m}\ket{s}_A\ket{0^n}_D.
\]
Apply controlled state-preparations on $D$:
if $s\neq r$ prepare $\ket{b_s}$ from $\ket{0^n}$, and if $s=r$ prepare $\ket{x}$.
This gives
\[
\ket{\Psi_1}
=\frac{1}{\sqrt{2^m}}\Big(\sum_{s\neq r}\ket{s}\ket{b_s}+\ket{r}\ket{x}\Big).
\]
Apply a controlled-$\tilde O$ conditioned on $r$:
\[
\ket{\Psi_2}
=\frac{1}{\sqrt{2^m}}\Big(\sum_{s\neq r}\ket{s}\ket{b_s}+\ket{r}\,\tilde O\ket{x}\Big).
\]
Finally apply $H^{\otimes m}$ on $A$. Using
\(
H^{\otimes m}\ket{s}
=2^{-m/2}\sum_{a}(-1)^{a\cdot s}\ket{a}
\)
(the inverse Hadamard transform),
we obtain
\[
\ket{\Psi_3}
=\frac{1}{2^m}\sum_{a}\ket{a}\Big(
\sum_{s\neq r}(-1)^{a\cdot s}\ket{b_s}
+(-1)^{a\cdot r}\tilde O\ket{x}
\Big).
\]
Define the (unnormalized) conditional data state
\[
\ket{v_a}:=\sum_{s\neq r}(-1)^{a\cdot s}\ket{b_s}+(-1)^{a\cdot r}\tilde O\ket{x}.
\]
Then the ancilla outcome probability is
\[
p_a=\Pr(A=a)=\frac{\|\ket{v_a}\|^2}{2^{2m}}.
\]
Because the $\{\ket{b_s}\}_{s\neq r}$ are orthonormal and $\|\tilde O\ket{x}\|=1$
(since $\tilde O$ is unitary), we have
\[
\begin{aligned}
\|v_a\|^2
&=\sum_{s\neq r}\|b_s\|^2 + \|\tilde O x\|^2
\end{aligned}
\]
\[
\begin{aligned}
&+2\,\Re[ \,\sum_{s\neq r}(-1)^{a\cdot s}(-1)^{a\cdot r}\bra{b_s}\tilde O\ket{x} \,]
\end{aligned}
\]
\[
\begin{aligned}
&=(2^m-1)+1
+2\sum_{s\neq r}(-1)^{a\cdot(s\oplus r)} r_s\\
&=2^m+2\sum_{s\neq r}(-1)^{a\cdot(s\oplus r)} r_s.
\end{aligned}
\]
Hence
\begin{equation}
p_a
=\frac{1}{2^m}
+\frac{1}{2^{2m-1}}
\sum_{s\neq r}(-1)^{a\cdot(s\oplus r)} r_s
\label{eq:multiprob}
\end{equation}
We can invert Eq.~\ref{eq:multiprob} as follows.
Let $\Delta_a:=p_a-2^{-m}$. Multiply  Eq.~\ref{eq:multiprob} by $(-1)^{a\cdot(t\oplus r)}$
and sum over all $a$:
\[
\sum_a(-1)^{a\cdot(t\oplus r)}\Delta_a
=\frac{1}{2^{2m-1}}\sum_{s\neq r} r_s
\sum_a (-1)^{a\cdot(s\oplus t)}.
\]
Using the identity
\(
\sum_a (-1)^{a\cdot(s\oplus t)} = 2^m\,\delta_{s,t}
\), we find that, for any $t\neq r$
\begin{equation}
r_t
=2^{m-1}\sum_a (-1)^{a\cdot(t\oplus r)}
\Big(p_a-\frac{1}{2^m}\Big)
.
\end{equation}
Thus measuring all ancillas provides the full probability vector $(p_a)$,
from which classical post-processing generates every $r_t$
simultaneously. Since the attribution gradients satisfy $g_t=2r_t$
(from Lemma~\ref{lemma:gradientcircuit}),
this yields $2^m-1$ gradient components in parallel using only $m$ ancillas.
\end{proof}

Increasing the number of ancillae will also increase the compilation overhead of the \sol{} circuit. This is especially pertinent in the near term, where circuit depth and gate count will limit the length of computations that can be run on hardware.  As such, we highlight that \sol{} 's flexibility regarding the number of ancilla qubits is highly desirable, as it allows users to fine-tune the use of quantum resources.

This aligns with a common design principle of the so-called Early Fault-Tolerant Quantum Computing (EFTQC) era: trade reduced circuit complexity and qubit count for increased circuit evaluations (shots)~\cite{modelingEFTQC}.
\section{\sol{}'s Methodology and Evaluation}
\label{sec:evaluation}

\subsection{Experimental Setup}

We first test our technique on a variety of image datasets, including Bars and Stripes~\cite{bowles2024better}, MNIST~\cite{lecun1998mnist}, NIST (similar to MNIST, but with reduced resolution: 8$\times$8), and FashionMNIST~\cite{xiao2017fashion}. For each image dataset, we construct binary classification tasks by randomly selecting two classes. 

We also test \sol{} on a synthetic quantum dataset based on the 1D transverse-field Ising model (TFIM) to emphasize its broad applicability to QML tasks beyond image classification. The TFIM Hamiltonian can be written as $H = -J \sum_{\expval{i,j}}\sigma^z_i - g\sum_i \sigma^x_i$. Here, $\expval{i,j}$ denotes a sum over only nearest neighbor sites on the chain. For our synthetic dataset, we set $J=1$ and generate random samples of $g$ pulled from a uniform distribution on the interval $[0,2]$. We focus on classifying two distinct ground-state phases: the disordered phase ($g > 1$) and the ordered phase ($g < 1$). We generate datapoints as the set of pairwise correlation functions between the first spin and all others in the chain, similar to \cite{franco_quantum_SHAP2026}. This gives us datapoints of the form:
\begin{align*}
    \vect{x} = \big( &\expval{\sigma^x_0 \sigma^x_1}, \expval{\sigma^x_0 \sigma^x_2}, \ldots, \expval{\sigma^x_0 \sigma^x_{L-1}}, \\
    &\expval{\sigma^y_0 \sigma^y_1}, \expval{\sigma^y_0 \sigma^y_2}, \ldots, \expval{\sigma^y_0 \sigma^y_{L-1}}, \\
    &\expval{\sigma^z_0 \sigma^z_1}, \expval{\sigma^z_0 \sigma^z_2}, \ldots, \expval{\sigma^z_0 \sigma^z_{L-1}},\big), 
\end{align*}
where $L$ is the length of the 1D chain. As such, successful classification relies on the model's ability to identify the phase based on spatial correlations among spin operators at each site.

\begin{figure}[t]
    \centering
    \subfigure[Raw Images]{\includegraphics[width=0.32\linewidth]{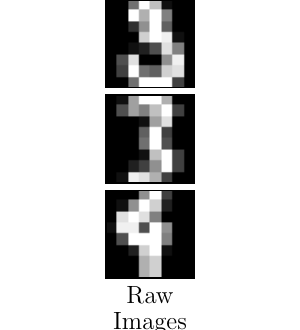}}
    \subfigure[Ideal Simulation]{\includegraphics[width=0.32\linewidth]{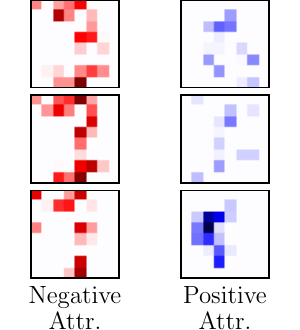}}
    \subfigure[Noisy Simulation]{\includegraphics[width=0.32\linewidth]{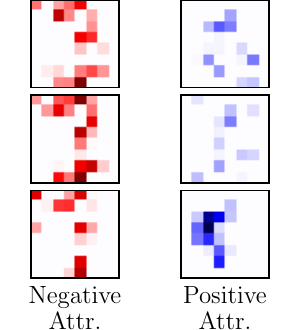}}
    \vspace{-3mm}
    \caption{Integrated gradients computed for three samples (a) of the NIST 3/4 classification task. In (b), simulation is performed assuming ideal hardware. In (c), a depolarizing error channel is added to the simulation, with error rate $10^{-4}$. Overall, we observe minimal degradation in the noisy hardware attribution scores compared to the ideal case.}
    \label{fig:hardware_noise}
    \vspace{-5mm}
\end{figure}

Training, inference, and gradient calculations are all performed in simulation, assuming ideal error-corrected hardware, except for a selection of results which undergo noisy simulation (Fig.~\ref{fig:hardware_noise}). As simulation is computationally prohibitive for larger systems, we focus on smaller, but representative, datasets to evaluate \sol{}. Our trained circuits are all constructed from a hardware-efficient ansatz consisting of alternating rows of single-qubit rotations and two-qubit CNOT gates. These layers are repeated multiple times to increase the number of model parameters. An example with 4 qubits is shown below:

\begin{center}
\begin{quantikz}[column sep=5mm,row sep=2mm,background color=red!20]
\qw & \gate[style=rot]{R_x(\theta_0)}& \gate[style=rot]{R_z(\theta_4)} & \ctrl{1} & \qw      & \qw      & \qw     \\
\qw & \gate[style=rot]{R_x(\theta_1)}& \gate[style=rot]{R_z(\theta_5)} & \gate[style=cnot]{X}  & \ctrl{1} & \qw      & \qw    \\
\qw & \gate[style=rot]{R_x(\theta_2)}& \gate[style=rot]{R_z(\theta_6)} & \qw      & \gate[style=cnot]{X}  & \ctrl{1} & \qw   \\
\qw & \gate[style=rot]{R_x(\theta_3)}& \gate[style=rot]{R_z(\theta_7)} & \qw      & \qw      & \gate[style=cnot]{X}   & \qw \\
\end{quantikz}
\end{center}
In such a circuit, the number of parameters is proportional to the number of qubits $\cross$ the number of layers. Generally, selecting a layer count between 1$\cross$ and 2$\cross$ times the number of qubits provides the best accuracy (as demonstrated by our selection for the number of layers in Table~\ref{tab:models}).

All simulation code is written in Python 3.12.1, using Qiskit 2.0.0~\cite{qiskit2024} and PennyLane 0.41.1~\cite{bergholm2022pennylaneautomaticdifferentiationhybrid}. Data preprocessing was performed with scikit-learn 1.6.1~\cite{scikit-learn}, and the optimization for training circuit parameters was performed using COBYLA~\cite{COBYLApowell1994direct}, as implemented in Scipy 1.15.1~\cite{2020SciPy-NMeth}. Due to the difficulty of training the angle-embedded model, we used a gradient descent optimizer implemented in PennyLane. Experiments are run on a local cluster comprising nodes with the AMD EPYC 7702P 64-core processor. We spawn virtual machines with 8 cores and 32 GB of memory.

\subsection{Model Architecture}

To focus on the general applicability of our technique, we choose to train relatively simple models (model properties are shown in Table~\ref{tab:models}) based on the hardware-efficient ansatz, which is composed of alternating layers of single-qubit rotation gates and two-qubit CNOT gates~\cite{cerezo2021variational}. Data is encoded in the system using an amplitude-encoding scheme, where the intensity of a pixel corresponds to the amplitude of a basis state. For these datasets, it is not generally true that each data point is normalized with $|\vect{x}| = 1$, meaning we can not directly encode them as quantum states $\ket{x} = \sum_i x_i \ket{b_i}$, but must first apply some transformation. The easiest of these is to simply divide each data point by its norm; however, we find empirically that this can cause issues during training, as the absolute value of a pixel's amplitude can change from image to image, even when the intensity is the same, due to images having differing levels of overall brightness. We instead use an encoding scheme with an overflow state. This overflow state allows us to encode the value of each pixel in a way that is consistent image-to-image, while maintaining the normalization condition of quantum states. In an $n$ qubit model, we scale $2^n - 1$ pixel values $x_i$ to be within $[0, (\frac{1}{2^{n}-1})^{\frac{1}{2}}]$. The remaining state, the overflow state, is then assigned the value $\smash{(1-\sum_i^{2^n-1}|x_i|^2)^\frac{1}{2}}$ so that the final norm of the state is $1$. Measurement is performed on a single qubit in the Z basis, i.e., $O$ in Eq.~\ref{eq:model} is the single qubit Z operator. While testing, we found improved performance with a nonlinear tanh activation applied to the circuit's output. All of our discussion from before still applies upon simple modification using the chain rule.\begin{figure*}[t]
    \vspace{1mm}
    \centering
    \subfigure[Raw Images]{\includegraphics[width=0.18\linewidth]{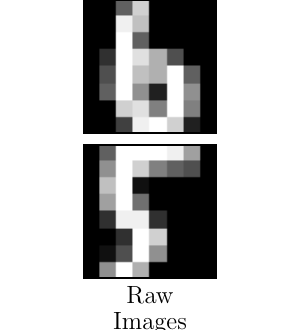}}
    \hspace{0.01\linewidth}
    \subfigure[10 shots]{\includegraphics[width=0.18\linewidth]{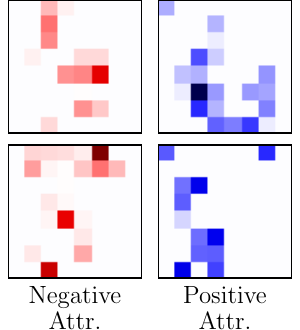}}
    \hspace{0.01\linewidth} % Space between subfigures
    \subfigure[100 shots]{\includegraphics[width=0.18\linewidth]{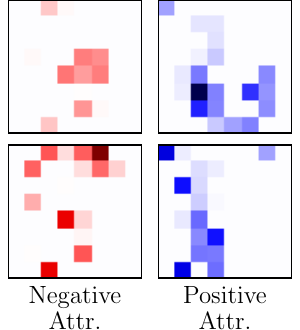}}
    \hspace{0.01\linewidth} % Space between subfigures
    \subfigure[500 shots]{\includegraphics[width=0.18\linewidth]{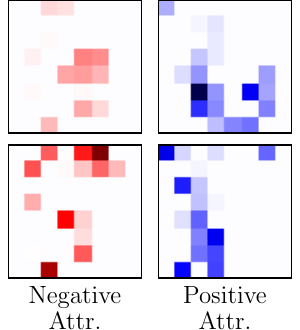}}
    \hspace{0.01\linewidth} % Space between subfigures
    \subfigure[Exact Sim.]{\includegraphics[width=0.18\linewidth]{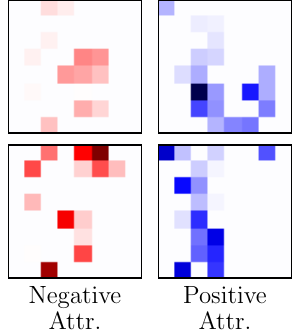}}
    \hspace{0.01\linewidth} % Space between subfigures
    \vspace{-3mm}
    \caption{Integrated gradients computed using various amounts of measurement shots (samples). In (b), (c), and (d), gradient components are computed using our circuit-based approach, using 10, 100, and 500 samples to estimate each component. Overall, we observe minimal degradation in the attribution scores compared to those obtained by exact simulation (e).}
    \label{fig:shot_noise}
    \vspace{-4mm}
\end{figure*}

\subsection{\sol{}'s Attribution Results}

We show attribution scores for a variety of samples across each dataset. These samples are chosen randomly for analysis. We use a blank image (all pixel values set to 0) as the baseline in all tests. While developing our technique, we also experimented with alternative baselines, including the average of all images in the training set. We observed similar behavior to the shown baseline. Positive and negative attributions refer to an input’s contribution toward the model's prediction, just as they do in the deep learning setting. Our model output is
$F(x;\theta) = \langle x | U^\dagger(\theta) O U(\theta) | x \rangle$. This model prediction also corresponds to the expected measurement outcome of the physical system, since it is the value used for the prediction. Thus, we attribute how redistributing amplitude (and hence probability mass via the Born rule) onto basis configuration $|b_k\rangle$ changes the observable $O$ along the input path. Positive IG means increasing the feature $x_k$ increases the class score; negative IG means the opposite.

In all plots, negative attributions are plotted in red, while positive attributions are plotted in blue. For visual clarity, attributions are normalized within each sample. Fig.~\ref{fig:NISTattributions} shows the integrated gradient outputs for a variety of samples from the NIST dataset. We see that background pixels have very little importance, as we might expect. We also see that the model has identified features that correspond to the target classes; an example being Fig.~\ref{fig:NISTattributions}(b), where we see negative attributions corresponding to the circular shape of the digit 3 toward the upper right, and positive attributions near the center left, corresponding to the angled shape of the digit 4. 

We see similar trends with MNIST (not shown for brevity), with the model attributing regions of each image to each class. In these larger examples, the attributions appear more mixed spatially. This is especially noticeable in the Dress/Shirt task, where a banding pattern forms, along with clusters of strong attributions at the center. We observe that in some cases, the model identifies distinctive features. One example is the Bag/Sandal task, where we observe high attribution along the straps, which are present only on bags and never on sandals. We also observe this in the Coat/Sandal task, where the upper area consistently receives negative attributions; this area is unlikely to contain any part of a sandal due to its low profile near that end of the shoe. We compare the attribution scores of a model using angle embedding with those of a model using amplitude embedding to see whether attributions differ across encoding schemes. Despite achieving very similar final accuracy scores (Table~\ref{tab:models}), we see markedly different attributions for the Bars and Stripes dataset in Fig.~\ref{fig:BarsAndStripesAttr}. 

\begin{figure*}[t]
    \vspace{1mm}
    \centering
    \includegraphics[width=0.99\linewidth]{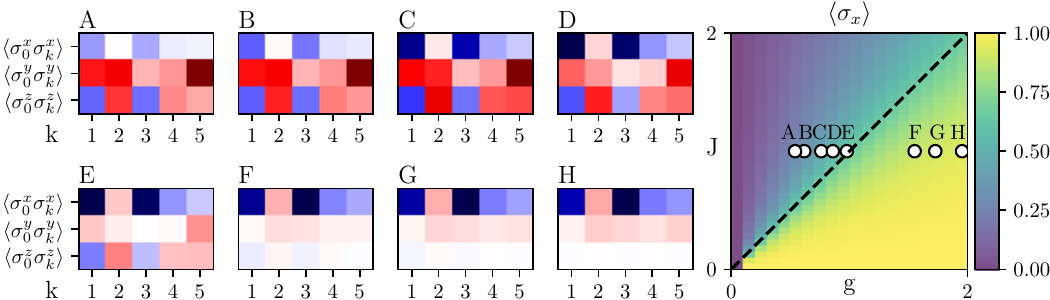}
    \vspace{-1mm}
    \caption{Attributions for samples drawn from the synthetic spin dataset. Positive (blue) attributions are associated with classification of the disordered state, while negative (red) attributions are associated with the ordered state. We see that in the strong field regime $(g > 1)$, strong positive attributions exist in the $\sigma^x$ correlation functions, while in the weak field regime $(g < 1)$, attributions emerge in the $y$ and $z$ components. We can see a transition in the attributions as we approach the phase boundary.}
    \label{fig:spins}
    \vspace{-2mm}
\end{figure*}

\section{Validation Against Null Model Attributions}

To validate the robustness and innovation of \sol{}'s attribution scores, we also compute attributions for null models having randomly generated parameters. Parameters are randomly sampled from either a uniform distribution on the interval $[0, \pi)$, a normal distribution $\mathcal{N}(0, \frac{\pi}{2}^2)$, and a heavy-tailed Student's t-distribution ($\nu = 2)$, Attributions are then computed and plotted for the same samples as used in Fig.~\ref{fig:NISTattributions}. Fig.~\ref{fig:nulluniform} shows the results for the uniform distributions (normal, and t-distributions display similar trends -- not shown). Across the various classification tasks, we fail to observe notable concentration or clustering of the attribution scores in any of the three distributions, unlike the trained case with \sol{}. For instance, the angular edge on the left side of digit four is only identified and attributed by \sol{}, while the three null attributions provide attribution scores all across the image.

\subsection{The Effect of Noise on \sol{}'s Performance}

To quantify the resource usage of \sol{}, we study the impact of reduced measurement shots on the ancilla qubit, thus introducing sampling noise into the attribution calculation. We compute attribution scores using 10, 100, and 500 shots to estimate each component $\Re[\bra{b_k}U^\dagger OU\ket{x}]$. Just as before, we repeat this for each component $k$ of the gradient, and use numerical integration to compute the IG attribution. We compare against exact simulation, which numerically computes all inner products directly from the state vector. The results of this are shown in Fig.~\ref{fig:shot_noise}. We observe that even with an extremely low shot count, the attribution scores computed are largely faithful to the numerically exact ones, with only small deviations appearing in some of the weaker attributions (those with a lower absolute value).

On the other hand, to probe the robustness of our technique in noisy operating conditions, we additionally simulate \sol{}'s performance under hardware noise representative of the early fault-tolerant quantum computing era. In this regime, logical operations remain subject to non-negligible residual error rates despite the presence of error correction, reflecting the expected constraints of near-term fault-tolerant devices where overheads preclude arbitrarily deep suppression~\cite{modelingEFTQC}. We model these effects using a standard depolarization channel, as implemented in Qiskit Aer~\cite{qiskit2024}:
\begin{equation}
    \mathcal{E}[\rho] = (1-\gamma)\rho + \gamma \, \text{Tr}[\rho] \, \frac{I}{2^n},
\end{equation} where $\rho$ is the density matrix of the system, $\gamma$ is the error rate and $n$ is the number of qubits. We focus on a regime of mild noise ($\gamma = 10^{-4}$) to simulate devices that are less noisy than current devices but not yet fully fault-tolerant. To keep our discussion general, we transpile all circuits to the basis $\{U,\, CNOT,\, CCX\}$, and assume full qubit connectivity. This choice of basis also makes transpiling our circuit straightforward, as we can simply attach an additional ancilla control to each gate in the ansatz. The controlled rotations are then transpiled further, while the CCX gates remain. We demonstrate integrated gradient calculations on the NIST 3/4 classification task using a single ancilla.

Notably, the resulting performance, shown in Fig.~\ref{fig:hardware_noise}, closely aligns with that observed in idealized, noiseless simulations: the generated scores deviate only mildly, preserving spatial distribution and relative intensity. These findings indicate that our method is well-suited to early fault-tolerant noise levels.

\subsection{Quantum Dataset Validation of \sol{}'s Performance}
Attributions for the TFIM spin dataset are presented in Fig.~\ref{fig:spins}. We assign class label 0 to the ordered phase $(g < 1)$ and class label 1 to the disordered phase $(g > 1)$; red (blue) attributions therefore correspond to the ordered (disordered) state. Across our testing samples, we observe strong positive attributions in the $\sigma^x$ correlation functions for the strong-field regime (samples F, G, and H). This agrees with physical intuition: within this regime, spins are largely aligned with the external field along the $x$ direction, and the magnitudes of the $y$ and $z$ components are minimized. These attributions contrast with the weak-field regime, where negative attributions emerge in the $y$ and $z$ components (samples A, B, and C). We can see a transition between these in the attributions as we approach the phase boundary (samples D and E).
\section{Related Work}

Several recent works have explored interpretability in QML, though none target input attribution directly, and none provide a general, hardware-compatible gradient-based solution as in \sol{}. Recent efforts in QML interpretability span model-agnostic techniques, gradient-based methods, and visualization tools. Pira et al.~\cite{pira2024interpretability}, Jahin et al.~\cite{jahin2023qamplifynet}, and Franco et al.~\cite{franco_quantum_SHAP2026} apply classical attribution methods like LIME and SHAP to QML models, while Heese et al.~\cite{heese2025explaining} use Shapley values to explain circuit components. These approaches rely on perturbation-based estimates and surrogate-based analysis, hence are not designed for execution on quantum hardware.

Gradient-based methods, such as QGrad-CAM~\cite{lin2024quantum}, demonstrate attribution in hybrid models using class activation maps, but are limited to specific architectures and do not generalize to amplitude encoding schemes. Visualization-driven efforts like QuantumEyes~\cite{ruan2023quantumeyes} and interpretable model designs~\cite{flamini2024towards, duneau2024scalable, flam2022learning, ran2023tensor} focus on circuit behavior or latent representations rather than input-level attribution and are limited in hardware compatibility (e.g., photonics or trapped ions). \textit{In contrast, \sol{} provides the first gradient-based input attribution method for QML models.} It supports amplitude encoding and enables scalable attribution via Hadamard test circuits and parallel gradient evaluation, making it broadly applicable across quantum models and devices.

%On the interpretability of quantum neural networks~\cite{pira2024interpretability}

%Towards interpretable quantum machine learning via single-photon quantum walks~\cite{flamini2024towards}

%Quantumeyes: Towards better interpretability of quantum circuits~\cite{ruan2023quantumeyes}

%Tensor networks for interpretable and efficient quantum-inspired machine learning~\cite{ran2023tensor}

%Quantum gradient class activation map for model interpretability~\cite{lin2024quantum}

%Learning interpretable representations of entanglement in quantum optics experiments using deep generative models~\cite{flam2022learning}

%Scalable and interpretable quantum natural language processing: an implementation on trapped ions~\cite{duneau2024scalable}

%QAmplifyNet: pushing the boundaries of supply chain backorder prediction using interpretable hybrid quantum-classical neural network~\cite{jahin2023qamplifynet}

%Explaining quantum circuits with shapley values: Towards explainable quantum machine learning~\cite{heese2025explaining}

\section{Conclusion}
\label{sec:conclusion}

We presented \sol{}, a unified framework for gradient-based feature attribution in quantum machine learning models to address the substantial challenge of interpreting them. As the first-of-its-kind quantum interpretability method, \sol{} operates on exponentially scaling amplitude-encoding schemes and is designed for execution on quantum hardware, enabling circuit-based gradient computations. We plan to extend \sol{} to generate parameter/layer attributions for QML models to determine their importance for the QML task, potentially leading to disagreements across multiple runs. By leveraging a Hadamard test–based construction and a multi-ancilla parallelization strategy, \sol{} enables scalable, implementation-agnostic input attribution with fidelity guarantees. %We also plan to extend \sol{} to support QML models with mid-circuit measurements and conditional gate operations, which are now becoming available on quantum hardware. Due to the effectiveness and unitary nature of QML, it is also of interest to explore unitary feature learning, or equivalently, learning from spherical features. 

\section{Acknowledgement}

This work was supported by Rice University, Santa Clara University, the Rice University George R. Brown School of Engineering and Computing, and the Rice University Department of Computer Science. This work was supported by the DOE Quantum Testbed Finder Award DE-SC0024301, the Ken Kennedy Institute, and Rice Quantum Initiative, which is part of the Smalley-Curl Institute. Hengrui Luo was supported by the U.S. Department of Energy under Contract DE-AC02-05CH11231 and the U.S. National Science Foundation NSF-DMS 2412403.

\balance

\bibliographystyle{plain}
\bibliography{main}

\end{document}